\documentstyle[preprint,version2,aps]{revtex}
\begin{document}
\draft
\begin{quote}
\raggedleft cond-mat/9709099
\\Submitted to {\em Phys. Rev. B}
\end{quote}
\begin{title}
\centering
Band Crossing and Novel Low-Energy Behaviour~~~~~~~~~~\\
 in a Mean Field Theory 
of a Three-Band Model on a $Cu$--$O$ lattice
\end{title}
\author{D. I. Golosov\footnotemark[1], A. E. Ruckenstein\footnotemark[2],
 and M. L. Horbach\footnotemark[3]}
\begin{instit}
Department of Physics, Rutgers University, Piscataway, NJ 08855-0849,
U.S.A.
\end{instit}
\footnotetext[1]{Present address: The James Franck Institute, The University of
Chicago, 5640 S. Ellis Avenue, Chicago, IL 60637, U. S. A.}
\footnotetext[2]{Also at: Institut f\"{u}r Theorie der Kondensierten
Materie, Universit\"{a}t Karlsruhe, 76132 Karlsruhe, Germany.}
\footnotetext[3]{Present address: Information Builders, 1250 Broadway, New York,
NY 10001, U. S. A.}
\begin{abstract}
We study correlation effects in a three-band
extended Hubbard model of ${\rm Cu}$--${\rm O}$ planes within the $1/N$
mean field  
approach, in the infinite $U$ limit. 
We investigate the emerging phase diagram and
discuss the low energy scales associated with each region.
With increasing direct overlap between oxygen orbitals, $t_{pp} >0$, the 
solution displays a band crossing which, for an extended range
of parameters, lies close to the Fermi level.
In turn this leads to the nearly nested character of the
Fermi surface and the resulting linear temperature dependence of
the quasi-particle relaxation rate for sufficiently large $T$.  
We also discuss the effect of 
band crossing on the optical conductivity and comment on the possible
experimental relevance of our findings.
\end{abstract}
\newpage

A number of versions of the extended three-band Hubbard models have 
been studied recently \cite{sirev,klr,auerbach,gkm,hicks}
with the expectation that they are
relevant to the physics of the high-$T_c$ cuprates.
The complexity of these models leads to a broad
spectrum of phenomena, from metal-to-insulator transitions,
charge and spin-density waves and superconductivity to 
more mundane single particle band-structure effects.
Most of the studies searched for novel low-energy excitations
which would explain the departures from Landau's
Fermi Liquid phenomenology  implicit in many of the experimental
results in the normal
state of the cuprates.

In the simplest discussions based on mean field theories of
strongly correlated Fermi systems such non-Fermi liquid behaviour
occurs above some small energy scale, $T_{coh}$. Already
within the three-band models there are a number of
different physical mechanisms resulting in the appearance of 
small energy scales, namely
the proximity of the chemical potential to a
van Hove singularity in the single-particle density of states~\cite{newns},
heavy Fermion behaviour~\cite{gkm} and 
the Brinkman-Rice metal-to-insulator transition~\cite{gkm,hicks,italians}, 
and the effects of 
strong anti-ferromagnetic spin fluctuations 
in the paramagnetic state~\cite{italians,simag}.

In this letter we focus on a band-crossing which arises upon hole
doping ($n_0 >1$) for values of
the $O-O$ hopping matrix elements 
larger than a filling-dependent critical value, $t_{pp} ^c > 0$.
In the vicinity of $t_{pp} =t_{pp} ^c$ the lowest ($Cu$-like) band is
only weakly dispersing along
the Brillouin zone boundary leading to a nearly
one dimensional ``extended" van Hove singularity in the density of
states (cf. Ref. ~\cite{abrikosov}).
In addition, once the doping is increased to a particular concentration, 
$n_0 = n_{cr}$, the band-crossing 
occurs {\em at} the Fermi level. (This rare
situation in the context of three dimensional 
band-metals is best exemplified by the self-intersecting
Fermi surface of graphite~\cite{graphite}.) Moreover, at $n_{cr}$,
the resulting Fermi surface (FS)
is perfectly nested with an incommensurate
nesting wave vector. A special feature of this situation is
that the nearly nested character of the Fermi surface
survives for a wide range of dopings around $n_{cr}$.
In particular, in the vicinity of $n_{cr}$ the resulting low-energy 
scale varies with doping slower ($T_{coh} \propto (n_0 -n_{cr})^2$)
than that expected from the saddle-point van Hove scenario 
($\propto -|n_0 -n_{cr}|/\ln |n_0 - n_{cr}|$).

We emphasize that in the experimentally interesting case of hole doping, 
the band-crossing only occurs for a particular sign
of $t_{pp}(>0)$. Under these conditions the van Hove singularity of
the density of states does not scan the Fermi level with increasing
doping and thus the van Hove mechanism discussed in ~\cite{newns}
cannot be realized. Since, generically, 
the crossing of two eigenvalues of a
hermitian operator requires the fine tuning 
of three real parameters~\cite{vonNeumann} band-crossing in two-dimensions
is typically avoided. Our situation is special as the mean field
Hamiltonian can be reduced to a real matrix and thus crossing
may occur through the tuning of only two parameters (e.g., the two
components of the momentum, $(k_x ,k_y )$). We note that avoided
crossing along
a line in $2$-dimensions (or at a point in $1$-D) is responsible for
the heavy Fermion regime of our mean field solution~\cite{kotliar1d}.
 
In the present paper, we concentrate on the model with $t_{pp}>0$. We note
that this choice disagrees with the situation which apparently takes place
in all the cuprates, where
the estimates extracted from LDA calculations
\cite{hybertsen,AndersenYBCO} imply that $t_{pp}$ is negative.
 In YBCO near optimal doping
the corresponding FS agrees qualitatively with
the ARPES results~\cite{shen}. Some questions however arose in the context
of BiSCO where two sheets appear in one of the earlier interpretations
of the photo-emission results~\cite{Dessau93} and there remains
controversy of whether 
the origin of the two sheets is a lattice superstructure effect, 
bilayer splitting or some other effect.
The main reason for studying a model with $t_{pp}>0$ is the fact that, as we
will see shortly, a rich array of many-body and single-particle phenomena
arise in a natural way. It is interesting, however, that in some cases one is
able to draw a parallel between our results and the experimental observations. 
If this model  has any relevance
it can only be so due to strong correlations
which are believed to be crucial to the physics of
the cuprates\cite{pwa} and are not properly accounted for within the LDA.
To exemplify this point, imagine
starting with a $t_{pp} =0$ 
tight binding model with a large
value of the Hubbard $U$ and deriving
an effective low-energy Hamiltonian by ``integrating out" 
charge fluctuations on the $Cu$-sites one obtains an effective
$O-O$ hopping amplitude which changes sign with increasing $U$.
In the $U\rightarrow \infty$ limit this yields $t_{pp} ^ {eff} >0$
(cf. Ref. \cite{gooding}). 

Our starting point is the extended three-band Anderson-Hubbard
Hamiltonian describing the interaction between 
the oxygen $p_x , p_y$ ($P_{x\sigma} ,P_{y\sigma}$)
and copper $d_{x^2 -y^2}$ ($d_{\sigma}$) orbitals. As 
depicted in Fig. 1, the Hamiltonian includes the direct 
hopping of electrons
between oxygen sites ($t_{pp}$) and a hybridization between copper and 
oxygen orbitals ($t_{pd}$). Apart from the
Hubbard repulsion at the copper sites -- here after taken as infinite --
we also take into account
the nearest neighbor Coulomb interaction, $V$, between copper and oxygen.
The infinite Hubbard repulsion is incorporated through the replacement
of the original copper orbitals operators, $d_{\sigma}$, by the projected
Fermion operators, $\tilde{d}_\sigma = d_{\sigma} (1-n_{-\sigma})$
which eliminate double occupancy at the copper sites; here $n_{\sigma '} =
d^{\dagger} _{\sigma '} d_{\sigma '}$ is the local number operator for
the orbital with spin $\sigma '$. Below we treat this Hamiltonian by
a simple mean field (Hartree-Fock) approximation: as explained in Refs. 
\cite{hicks,aer88},  
the factorization of the equations of motion requires a generalization
of the usual Wick's theorem.
The resulting mean field equations can be adequately described by
the quasi-particle Hamiltonian,
\begin{eqnarray}
{\cal H} _{MF}  = && -2i t'_{pd} \sum _{\vec{k} ,\sigma}
\tilde{d} ^{\dagger} _{\sigma} (\vec{k}) 
[\sin \frac{k_x}{2} P_{x\sigma} (\vec{k}) - 
\sin \frac{k_y}{2} P_{y\sigma} (\vec{k})] -4t_{pp}\sum _{\vec{k} ,\sigma}
P^\dagger_{x\sigma}(\vec{k})P_{y\sigma}(\vec{k})
\sin \frac{k_x}{2} \sin \frac{k_y}{2} + h.c. \nonumber \\
&&+ (\epsilon '_{d} -\mu )\sum _{\vec{k} ,\sigma} \tilde{d} ^{\dagger}
_\sigma (\vec{k}) \tilde{d} _{\sigma} (\vec{k}) +
(\epsilon '_{p} -\mu ) \sum _{\vec{k} ,\sigma} 
[P^{\dagger} _{x\sigma} (\vec{k})
P_{x\sigma} (\vec{k}) + 
P^{\dagger} _{y\sigma} (\vec{k})
P_{y\sigma} (\vec{k})] \,.
\label{eq:mfhamilt}
\end{eqnarray}
Here $t' _{pd} = \sqrt{1-n_d} (t_{pd} +V\lambda /8t_{pd})$, 
$\epsilon ' _d = \epsilon _d + \lambda + 4 n_p V$ and
$\epsilon ' _p = \epsilon _p +2n_d V$ are renormalized values of the   
hybridization amplitude, and local energy levels of the copper and oxygen
orbitals, $t_{pd}, \epsilon _d ,\epsilon _p$, respectively;
given $\Delta ^{(0)} = \epsilon _p - \epsilon _d$ and
the total filling, $n_0 = n_d + 2n_p$,
the values of the local copper ($n_d$) and oxygen ($n_p$)
occupancy,
the energy shift, $\lambda$, and the chemical potential, $\mu$,
are to be determined self-consistently from the mean field equations (MFE),
\begin{eqnarray}
n_0&= & \frac{2}{N} \sum_{\vec{k}} \sum_{i=1}^3
n_F \left( \epsilon_i(\vec{k}) - \tilde{\mu} \right)\,,
\label{eq:mfen0} \\
n_d & = & \frac{2}{N} \sum_{\vec{k}} \sum_{i=1}^3 n_F \left(
\epsilon_i(\vec{k}) - \tilde{\mu} \right)
\frac{\left(\epsilon_i(\vec{k})- \Delta\right)^2 -16 t_{pp}^2 \sin^2 \frac
{k_x}{2}\,\sin^2 \frac {k_y}{2} } {\prod_{j \neq i} \left
(\epsilon_i(\vec{k}) - \epsilon_j(\vec{k}) \right)}\,,
\label{eq:mfend} \\
\lambda &=& - \frac{8}{N} \sqrt{\frac{2}{1-n_d}} t_{pd} t'_{pd} 
\sum_{\vec{k}} \sum_{i=1}^3 n_F 
\left( \epsilon_i(\vec{k}) - \tilde{\mu} \right) 
\frac {4 t_{pp} \sin^2 \frac {k_y}{2} + \epsilon_i(\vec{k})- \Delta}  
{\prod_{j \neq i} \left 
(\epsilon_i(\vec{k}) - 
\epsilon_j(\vec{k}) \right)}\,\sin^2 \frac {k_x}{2}\, 
\label{eq:mfedelta} \end{eqnarray}
where $\tilde{\mu} = \mu -\epsilon ' _d$, $\Delta = \epsilon ' _p -
\epsilon' _ d$, N is the number of copper sites, 
and $n_F (x) = [\exp \beta x +1] ^{-1}$ is 
the Fermi-Dirac distribution function.
Finally, the quasi-particle energies, $\omega _i (\vec{k})=
\epsilon ' _d + \epsilon _i (\vec{k})$, $i=1 \div 3$, should be determined from
the cubic secular equation for $\epsilon _i (\vec{k})$, namely,
\begin{eqnarray}
\epsilon^3-2 \Delta \,\epsilon^2& +& \left [ \Delta^2 - 4 t^{\prime 2} _{pd} 
(\sin ^2 \frac{k_x}{2} +\sin^2 \frac {k_y}{2}) -16 t^2_{pp}\sin^2 \frac
{k_x}{2} \sin^2 \frac {k_y}{2} \right] \epsilon + \nonumber \\
&&+4\Delta\,t^{\prime 2} _{pd}
(\sin^2 \frac {k_x}{2} +\sin^2 \frac {k_y}{2}) - 32 t^{\prime 2}_{pd}
t_{pp}\sin^2 \frac {k_x}{2} \sin^2 \frac {k_y}{2} = 0 .
\label{eq:secular}
\end{eqnarray}

Equations (~\ref{eq:mfen0}--~\ref{eq:secular}) define the
filling-dependent bandstructure. 
The no-double occupancy constraint manifests itself through two important
effects:
(i) the upward shift of the copper-like band and (ii) 
its associated band narrowing. The first occurs
as a result of the no double occupancy
constraint which forces the copper component of the lower band to lie within
a hybridization width of the chemical potential; while the band narrowing
is due to the $\sqrt{1-n_d}$ factor multiplying $t_{pd}$
and reflects the fact that coherent hybridization involves
rare charge fluctuations at the copper sites.
The interplay between these correlation effects and the band structure 
leads to the rich phase diagram shown in Figs. 2.
To summarize the possible behaviours imagine starting with
the bare copper level below the oxygen level, $\Delta ^{(0)} >0$
and increasing the filling from $n_0 =0$. We will consider 
the $\Delta ^{(0)}$ vs. $n_0$ plane for two qualitatively
different cases, $t_{pp} <t_0$ and $t_{pp} >t_0$, where
$t_0 = \sqrt{\alpha _1} |t_{pd}| + \alpha _2 V$ with the two constants
given by $\alpha _1 =(4\sqrt{2} -5)/{\pi} \approx .21$ and
$\alpha _2 =[\pi -4(\sqrt{2} -1)]/4\pi \approx .12$. 
In each case we single
out three $\Delta ^{(0)} = const$ cuts, denoted by $A, B,$ and $C$ in
the figures.
For $t_{pp} < t_0$ and for small values of $\Delta ^{(0)}$ (cut A in 
Fig. 2a) one starts with conventional metallic behaviour for $n_0 <<1$.
Below half filling 
one then scans the the chemical potential
through the van Hove singularity (saddle point) 
switching from hole-like to electron-like
FS.  
The situation at $n_0<1$
is reminiscent of what would happen at $n_0>1$ if we reverse the
sign of the oxygen-oxygen hopping. The low energy scale is provided by
the difference between the Fermi level and the saddle point energy
(see fig.3), and although in the case of the Fermi energy equal to
$\epsilon_{vHs}$ one does get the linear temperature dependence of the
quasiparticle relaxation rate \cite{newns}, the corresponding crossover
temperature $T_*$ 
above which the relaxation rate becomes linear in $T$
is in this case expected to scale as
$(n_0-n_{vHs})/\ln|n_0-n_{vHs}|$ \cite{gopalan}, where $n_{vHs}$ is
the value of filling corresponding to $\tilde{\mu}=\epsilon_{vHs}$.
The physical consequences of the presence of a small energy scale
$\tilde{\mu}-\epsilon_{vHs}$ have been discussed in the
literature in great detail \cite{vhreview}.

With further increasing $n_0$ the chemical potential scans through another
van Hove singularity associated with the bottom of the lower oxygen-like
band, which, at $T=0$, causes a negative jump
in the compressibility,
$dn_0/d\mu$. Above this doping concentration, $dn_0/d\mu <0$,
and the system is thermodynamically unstable with
respect to phase separation.
Within the region of instability, the absolute value of the (negative)
compressibility increases and, at some point, diverges.
The compressibility then changes sign and the system enters a
normal metallic phase with large and rapidly decreasing value of
the compressibility.

The shape of the unstable
region is highly sensitive with respect to the parameters of the
Hamiltonian. The phase separation would be circumvented altogether
in the presence of a real long-ranged Coulomb interaction.
Beyond half-filling the bottoms of the two lowest 
bands coincide (``tangency") beyond which a crossing of
the $Cu$- and lower $O$-like bands (conic point)
emerges. At sufficiently large filling
the chemical potential scans through the band crossing energy.
Along cut B in Fig. 2a a new feature arises at $n_0 =1$ 
(between the two van Hove
singularities), namely the Brinkman-Rice 
metal-to-insulator transition (BRT) where the copper-like
band becomes completely flat (see below). Finally,
along cut C, in addition to the BRT, a heavy fermion regime with
exponentially small $1 -n_d$ and a conic band crossing point below the
chemical potential appears immediately above half filling,
extends over a finite range of dopings and then
terminates via a smooth crossover into a conventional metal. 
In the case of cut C, the phase separation region is located above
this crossover.
In the metallic phase, the conic point eventually crosses the 
chemical potential.
For $t_{pp} > t_0$ both van Hove singularities and the
tangency appear already below half filling. The only other 
qualitative difference from the $t_{pp} < t_0$ case is the possible appearance
of reentrant heavy fermion behaviour below the BRT critical value
of $\Delta ^{(0)}$ (see cut B in Fig. 2b).
Numerical calculations show that the features of the phase diagram
persist at finite temperature, with the value of $\Delta ^{(0)}$
decreasing as temperature increases.
We are now in position to discuss these features in more detail.

The mean-field bandstructure is characterized by three energy scales,
namely the bandwidth of the lower band, the difference $\tilde{\mu} -
\epsilon_{vHs}$ between the Fermi level and the van Hove singularity
(corresponding to the saddle point in the lowest band dispersion)
and the difference between the Fermi level and the energy
$\epsilon_{cr}= -t_{pd}^{\prime\,2}/t_{pp}$ of the band crossing
point.
The values of these three quantities, computed along the three cuts A,
B, and C of fig. 2a are plotted in fig. 3. One can see, that indeed the
lowest energy scale in the metallic phase above half filling over an
extended range of dopings is
provided by $\tilde{\mu}-\epsilon_{cr}$, which vanishes at
some particular filling $n_{cr}(\Delta ^{(0)})$ and at small values of
$|n_0-n_{cr}|\ll 1$ depends linearly on filling,
$\tilde{\mu}-\epsilon_{cr} \propto n_{cr}-n_0$.

The FSs that emerge when the Fermi level lies close to $\epsilon_{cr}$
are shown  in Fig. 4  ~\cite{strontium}.
It is a striking feature of the present model
that when the band crossing occurs exactly at the Fermi level, the FS
is formed by the straight lines $k_x=k_{cr}$ and $k_y=k_{cr}$ (with 
$k_{cr}=\pi-\pi n_{cr}/4$), and is therefore perfectly nested along
the coordinate axes. 
It is well-known \cite{Ruvalds} that in such a
situation the quasiparticle relaxation rate becomes linear in
temperature. One can therefore expect that as the value of filling
approaches $n_{cr}$, the quasiparticle relaxation rate undergoes a
crossover to the linear temperature dependence; this is also the case
when, at some fixed value of $n_0$, temperature increases beyond
certain crossover value $T_*$.

The simplest way to exemplify this behaviour is to replace
the nested parts of the FS by two
nearly straight segments of length $L$ and curvature radius $R \gg L$,
also assuming that the velocities of quasiparticles 
on these segments are equal
to $v$ and antiparallel to each other. One can then estimate the
contribution of these segments to the off-shell decay rate due to a
weak interparticle contact interaction, $U\delta (\vec{r} - \vec{r}\,')$.
For $\omega > v L^2/R$,
and when momentum
vector
$\vec{p}$ lies at the middle of the straight segment of the FS,
the decay rate is linear in frequency, 
${\rm Im} \Sigma(\omega,\vec{p}) \approx (3 U^2 L^2 \omega)/(16 \pi^3
v^2)$ whereas at very low frequencies, $\omega \ll v L^2/R$, one gets
\begin{equation}
{\rm Im} \Sigma(\omega,\vec{p}) \approx \frac {RU^2}{(2\pi v)^3}
\left[ 2 \omega^2 \ln \frac {v L^2}{\omega R} - \frac {1}{3}\omega^2
\right]
\,.
\label{eq:taunormal}
\end{equation}
The finite-temperature
on-shell quasiparticle relaxation rate can be estimated by
substituting $T \rightarrow \omega$ in these expressions.
The quantity  $L^2/R$ measures the deviation of the
nearly-flat segments of the Fermi surface from the tangent straight
lines and provides a new low-energy scale, $v L^2/R$.
As expected ~\cite{Ruvalds}, for values of $\omega ,T >> v L^2/R$
${\rm Im}\Sigma$ is
proportional to $x = {\rm max} \{ \omega (>0) ,T \} $,
whereas at low energies the behaviour reduces to the 2D Fermi Liquid
result ${\rm Im}\Sigma \propto -x^2 {\rm ln} x$~\cite{Hodges}. 

For our mean field band structure 
$R\propto (\tilde{\mu}-\epsilon_{cr})^{-2}$ in the limit of 
$|\tilde{\mu}-\epsilon_{cr}/\tilde{\mu} | <<1$ and $L\sim \pi$.
Since $\tilde{\mu}-\epsilon_{cr} \propto n_{cr}-n_0$, it follows that
the crossover scale $T^* = v L^2 /R \propto (n_0-n_{cr})^2$.
As a result,
the linearity of the relaxation rate survives over a wider range of
dopings than in the van Hove scenario for which
$T^* _{vh} \propto |(n_0 - n_{cr})/ln (n_0 - n_{cr} )|$.

Let us now take a closer look of the band crossing point.
Expanding the dispersion law about this point leads to the following
two branches,
\begin{equation}
\epsilon_{1,2}(\vec{k}) = A(\vec{k}) \mp \sqrt{ B(\vec{k})(k_x +k_y-2
k_{cr})^2- C(\vec{k})(k_x-k_y)^2}\,,
\label{branches}
\end{equation}
where $A,B$ and $C$ are smooth functions of momentum $\vec{k}$.
Note that the square-root behaviour in (\ref{branches}) is
associated with the non-analytic behaviour of the matrix elements
of the unitary transformation which diagonalizes the mean field
Hamiltonian; more precisely,
the limiting values of these coefficients at $\vec{k} \rightarrow
\vec{k}_{cr}$ depend on the direction of approach.
This affects, for example, the interband matrix element of
the quasi-particle position operator which diverges at the
crossing point as
$\vec{r}_{12}(\vec{k})\propto
(\epsilon_1(\vec{k})-\epsilon_2(\vec{k}))^{-1}$.
On the other hand, the corresponding interband matrix element 
of the velocity operator
$\vec{v}_{12}=(d\vec{r} /dt)_{12} =
i(\epsilon_1-\epsilon_2)\vec{r}_{12}$
remains finite at $\vec{k} \rightarrow \vec{k}_{cr}$ but its limiting 
values still depend on the direction of approach. In turn, the
leading interband contribution to the optical conductivity (ignoring
all quasi-particle interactions)~\cite{Musik},
\begin{equation}
{\rm Re} \sigma ^{12}_{\alpha\beta}=\frac{e^2}{2\pi \omega} \int
v_{12}^\alpha(\vec{k}) v_{21}^\beta(\vec{k}) \left[
 n_F(\epsilon_1(\vec{k}))- n_F(\epsilon_2(\vec{k}))\right]
 \delta(\epsilon_1(\vec{k}) - \epsilon_2(\vec{k})+\omega) d^2k\,,
\label{eq:optical}
\end{equation}
(the indices $\alpha,\beta$ take values $x$ or $y$) displays
an anomaly associated with the peculiar behaviour of $\vec{v} _{12}$
in the presence of band crossing.
The integration in (~\ref{eq:optical}) leads to the
result sketched in fig. 5. 
Note the presence of two square-root singularities in the frequency
derivative of the conductivity: one at the
threshold frequency 
\begin{equation}
\omega_t=2\frac{t^2_{pd}+t_{pp}\Delta}{3t^2_{pd}+2t_{pp}\Delta}
|\mu-\epsilon_{cr}|\,,
\label{eq:threshold}
\end{equation}
below which ${\rm Re} \sigma ^{12}_{\alpha\beta}(\omega)$ is equal to
zero, and another at
$\omega^\prime_t=\omega_t \cdot (3t^2_{pd}+2t_{pp}\Delta)/t^2_{pd}$.
These singularities originate from the tangency between the ellipse
$\epsilon_2(\vec{k})=\epsilon_1(\vec{k})+\omega$ and the two borders
of the stripe $\tilde{\mu}<\epsilon_1(\vec{k})<\tilde{\mu}+\omega$.
At larger frequencies, $\omega \gg \omega_t,\omega_t^\prime$ (but $\omega$
still much smaller than the bandwidths),
${\rm Re} \sigma^{12}(\omega)$ approaches a constant
value. Finally, if the band band crossing occurs precisely
at the Fermi level, the independence of ${\rm Re} \sigma _{12} (\omega )$
on frequency survives down to $\omega = 0$. It is amusing to note
the resemblance of the threshold behaviour in Fig. 5 with the
``mid-infrared peak" observed in optical conductivity
experiments~\cite{infrared}. Within this scenario the weak
doping dependence of the ``peak" position could only be accounted
for if the system is sufficiently far below the critical
filling, $n_{cr}$.

It is easy to see that the value of the threshold frequency and the
nature of singularities at $\omega=\omega_t, \omega_t^\prime$ is in
fact unaffected by the RPA corrections. The latter, in principle,
might add excitonic features to the profile of  ${\rm Re}
\sigma_{\alpha\beta}(\omega)$  (delta-functional peaks below the
threshold or 
Lorentzian peaks above the threshold). We have checked, however, that
excitonic bound states do not occur for 
physically relevant interaction strengths
($V < W^2 /{\omega _t}$, where $W$ is  the conduction electron
bandwidth). Also, the effects of resonances above
threshold is small for $\omega \sim \omega _t$.
Thus, at frequencies of the order of $\omega _t$
RPA modifies the rigid-band
result by a factor of order unity.

We close by noting that when the Fermi level lies 
in the vicinity of a band crossing point,
one should be able to observe the phenomenon of
{\em magnetic breakdown}
\cite{Slutzkin}, namely,
the restructuring of the FS that takes
place at sufficiently high magnetic fields through the tunneling
of electrons between different FS sheets. This phenomenon and its consequences
for magneto-transport in weak fields in the presence of
a conic point will be described elsewhere~\cite{mb}.

We take our pleasure in thanking  
R. J. Gooding, M. I. Kaganov,  and
I. E. Trofimov for 
stimulating and enlightening discussions. This work was supported in part by 
ONR Grant \# N00014-92-J-1378.

\figure{The arrangement of copper $d$-orbitals and oxygen $p$-orbitals
in the ${\rm CuO}_2$ plane, with signs ``+'' and ``-'' accounting for
the phase factors in the atomic wave functions, which result in the
alternating signs of hopping terms in the tight-binding 
Hamiltonian.}
\figure{The phase diagrams for $t_{pd}=1.3$ eV, $t_{pp}=0.65$ eV, and 
$V=1.25$ eV (a), and for  $t_{pd}=1.3$ eV, $t_{pp}=0.8$ eV, and 
$V=0.5$ eV (b). Notice the difference in the shapes of the heavy
fermion (HF) region. The solid bold lines represent BRT, the dashed
bold lines - a smooth crossover between HF and normal metal
behaviour. The dotted line in fig. 3b corresponds to analytical
result for the HF to normal metal crossover, obtained in the limit of
small $n_0-1$.}
\figure{The bandstructure energy scales: the bandwidth of the lowest
band (solid lines), $\tilde{\mu}-\epsilon_{cr}$ (dashed lines), and
$\tilde{\mu}-\epsilon_{vHs}$ (dotted lines), computed along the three
cuts A, B, and C in fig. 2a.}
\figure{The Fermi surfaces for $n_0=1.23$ (dotted lines), $n_0=1.46$
($n_0 \approx n_{cr}$, solid lines), and $n_0=1.64$ (dashed
lines). The bare parameters of the system are $t_{pd}=1.3$ eV,
$t_{pp}=0.65$ eV,  $V=1.25$ eV , and $\Delta^{(0)}=3.26$ eV (cut B in
fig. 2a), and the number 0, 1 and 2 indicate the number of filled
bands in each region of the Brillouin zone.}
\figure{The rigid-band estimate for the interband term in the optical
$(ab)$-plane 
conductivity  (${\rm ohm}^{-1}{\rm cm}^{-1}$) versus frequency (eV) 
at $n_0=1.44$, $t_{pd}=1.3$ eV,
$t_{pp}=0.65$ eV,  $V=1.25$ eV , and $\Delta^{(0)}=3.26$ eV.
We used the value of 12 ${\rm \AA}$ for the interplane spacing.}
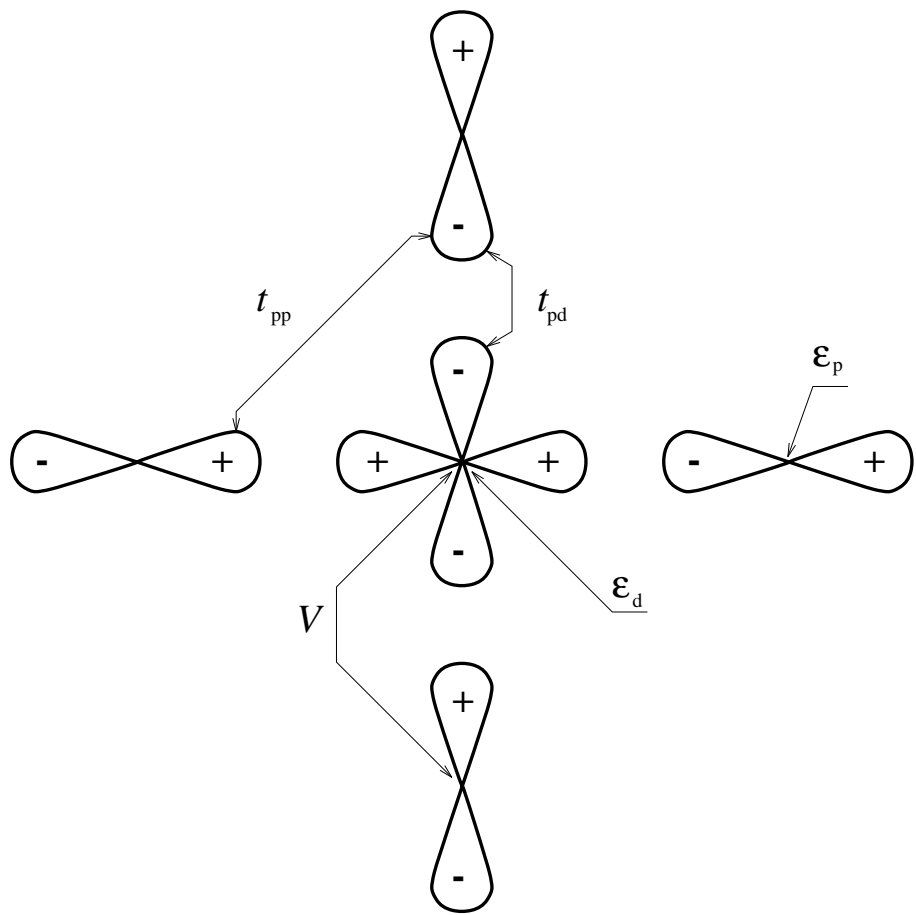
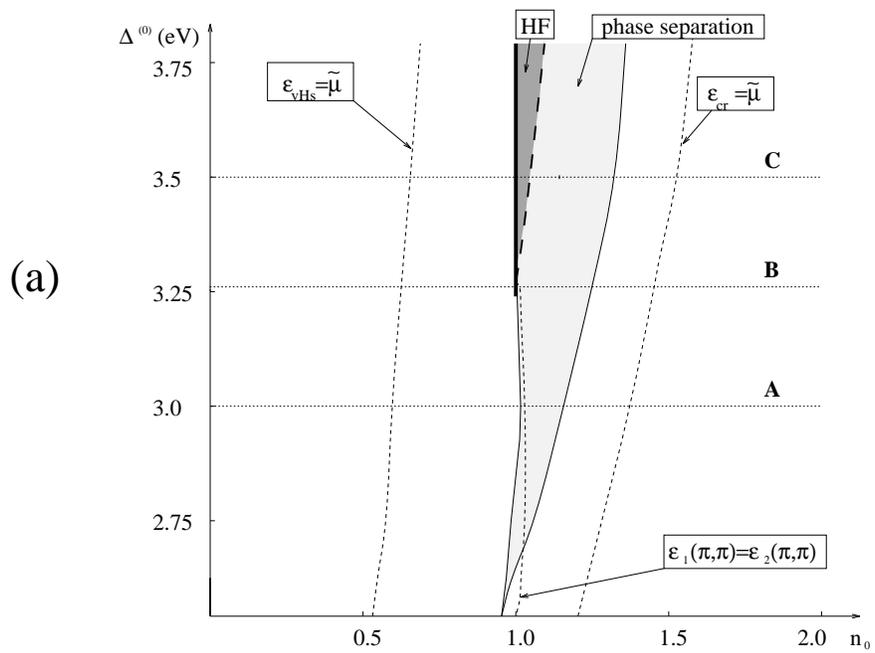
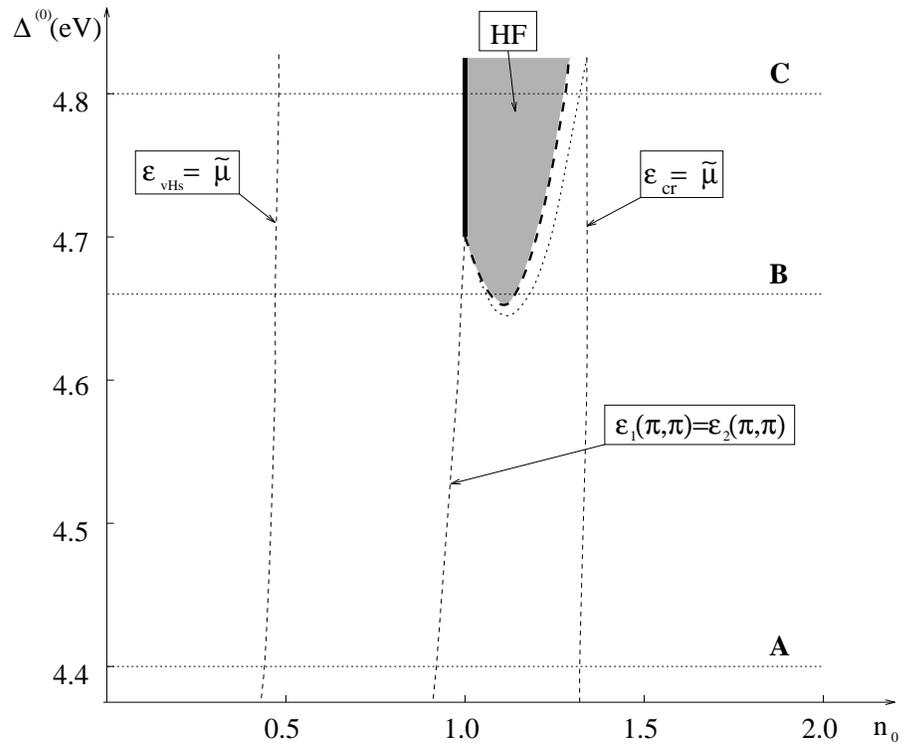
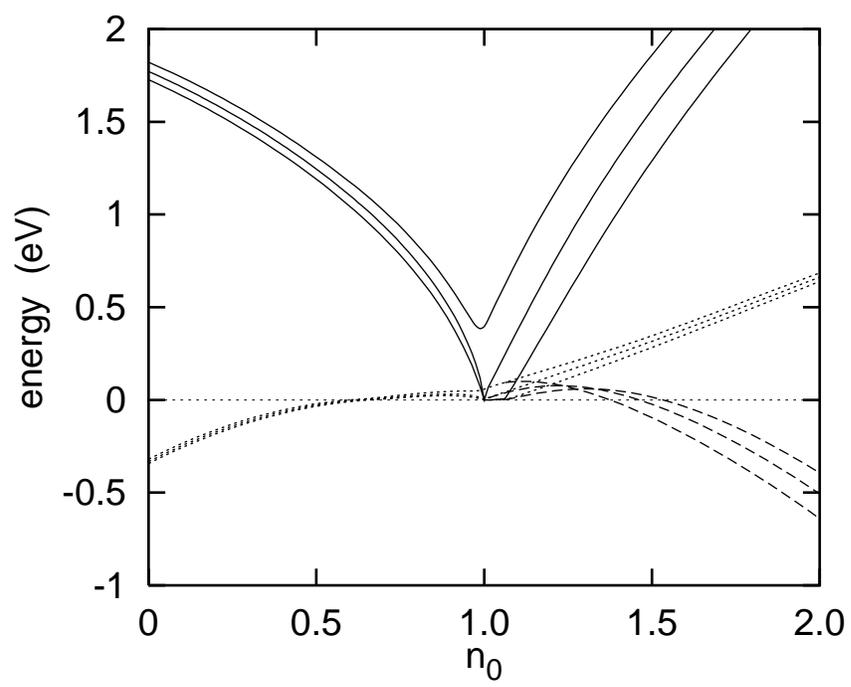
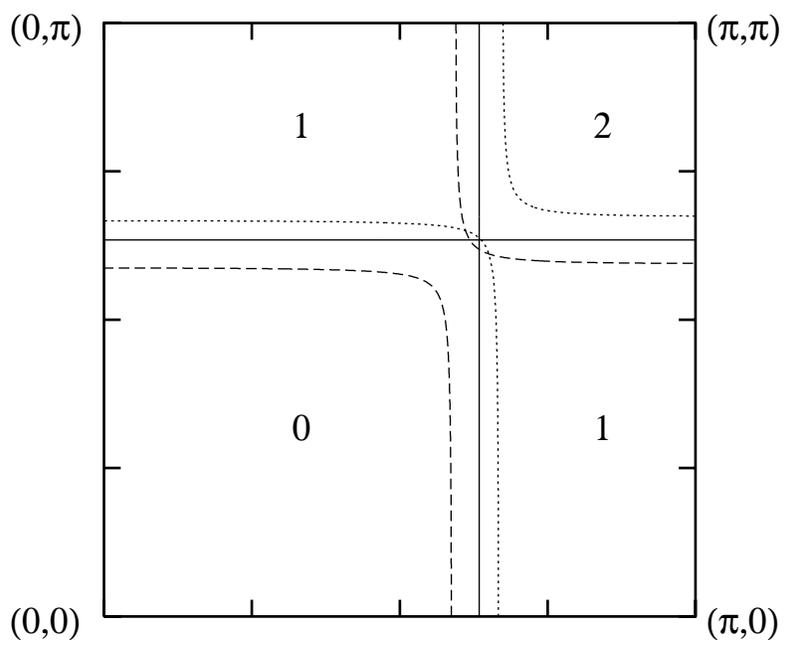
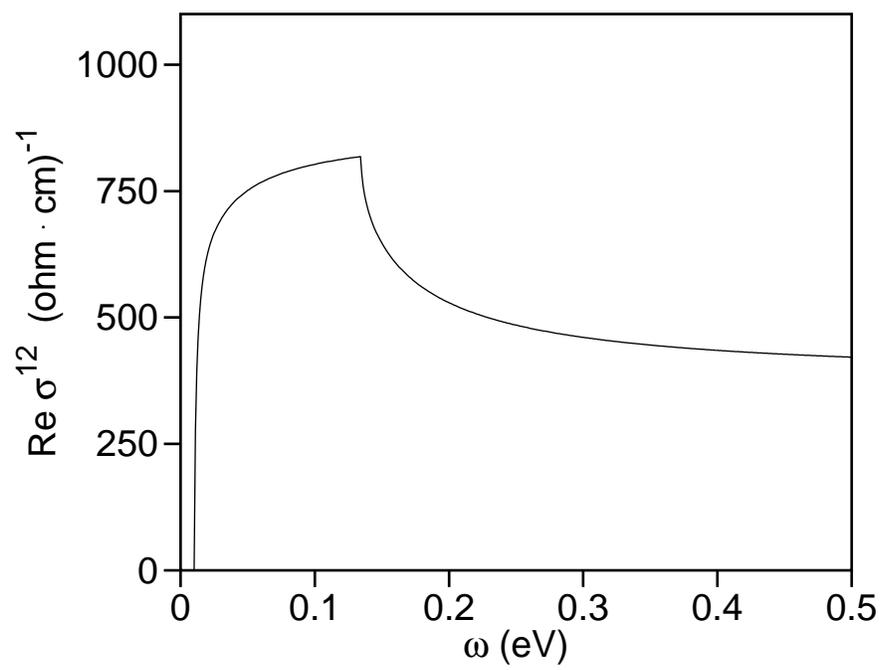
\end{document}